\documentclass{aa}
\usepackage{graphics}
\newcommand{\logteff} {\mbox{${\rm log}\, T_{\rm eff}$}}
\newcommand{\kmprs}  {\mbox{\rm\,km\,s$^{-1}$}}

\newcommand{\feh} {\mbox{\rm [Fe/H]}}
\newcommand{\nsis} {\log \varepsilon_{\odot} ({\rm Si})}
\newcommand{\nss} {\log \varepsilon_{\odot} ({\rm S})}
\newcommand{\nfes} {\log \varepsilon_{\odot} ({\rm Fe})}
\newcommand{\sh} {\mbox{\rm [S/H]}}
\newcommand{\ssi} {\mbox{\rm [S/Si]}}
\newcommand{\sih} {\mbox{\rm [Si/H]}}
\newcommand{\teff}  {\mbox{$T_{\rm eff}$}}
\newcommand{\logg}  {\mbox{{\rm log} $g$}}
\newcommand{\Ulsr} {\mbox{$U_{\rm LSR}$}}
\newcommand{\Vlsr} {\mbox{$V_{\rm LSR}$}}
\newcommand{\Wlsr} {\mbox{$W_{\rm LSR}$}}
\newcommand{\afe}  {\mbox{${\rm [\alpha/Fe]}$}}  
\newcommand{\XFe} {\mbox{\rm [X/Fe]}}
\newcommand{\OFe} {\mbox{\rm [O/Fe]}}
\newcommand{\SFe} {\mbox{\rm [S/Fe]}}
\newcommand{\SiFe} {\mbox{\rm [Si/Fe]}}
\newcommand{\MgFe} {\mbox{\rm [Mg/Fe]}}
\newcommand{\CaFe} {\mbox{\rm [Ca/Fe]}}
\newcommand{\ffe}  {\mbox{${\rm [\frac{Fe}{H}]}$}}
\newcommand{\sfe}  {\mbox{${\rm [\frac{S}{Fe}]}$}}  
\newcommand{\sife} {\mbox{${\rm [\frac{Si}{Fe}]}$}}
\newcommand{\Oone} {\ion{O}{i}}
\newcommand{\Sone} {\ion{S}{i}}
\newcommand{\Feone} {\ion{Fe}{i}}
\newcommand{\Sione} {\ion{Si}{i}}
\newcommand{\Fetwo} {\ion{Fe}{ii}} 

\begin{document}
\title{Sulphur abundances in disk stars: a correlation with silicon
\thanks{Based on observations carried out at National Astronomical
  Observatories (Xinglong, P.R. China)}}
\author{Y.Q.~Chen \inst{1,2} \and P.E.~Nissen \inst{2} \and G.~Zhao \inst{1}
\and M.~Asplund \inst{3,4}}
\offprints{P.E.~Nissen (pen@ifa.au.dk)}
\institute{
National Astronomical Observatories, Chinese Academy of Sciences,
   Beijing 100012, P.R. China 
\and 
Institute of Physics and Astronomy, University of Aarhus, DK--8000
Aarhus C, Denmark 
\and Uppsala Astronomical Observatory, Box 515, SE-751\,20, Sweden
\and
Present address: Research School of Astronomy and Astrophysics,
Australian National University, Mount Stromlo Observatory,
Cotter Road, Weston, ACT 2611, Australia}
\date{Received / Accepted }

\abstract{We have performed new determinations of sulphur
and silicon abundances for a sample of
26 disk stars based on high-resolution, high 
signal-to-noise spectra. The results indicate a
solar $\SFe$ for $\feh >-0.3$, below which
$\SFe$ increases to $\sim$0.25 dex at $\feh =-1.0$.
We find that there is a good correlation 
between $\sh$ and $\sih$, indicating the same 
nucleosynthetic origin of the two elements. It seems that
the ratio of sulphur to silicon does not depend 
on metallicity for $\feh > -1.0$.
The implications of these results on models for the nucleosynthesis
of $\alpha$-capture elements and the chemical evolution of the
Galaxy are discussed.
}

\maketitle

\section{Introduction}
The trend of $\OFe$ vs. $\feh$ has recently been paid much attention
because works by different authors yield very dissimilar results.
Specifically, Israelian et al. (\cite{Israelian98}, \cite{Israelian01a}) 
and Boesgaard (\cite{Boesgaard99}) derived oxygen 
abundances from OH UV lines and the infrared $\Oone$ triplet and found a linear rise of $\OFe$ to  1.0
dex when the metallicity decreases to $\feh =-3.0$ dex. 
Some doubts have, however, recently been cast on the correctness
of these results from calculations based on the new 
generation of 3D hydrodynamical 
model atmospheres (Asplund \& Garc\'{\i}a P{\'e}rez \cite{Asplund01}). Other
studies (e.g. Nissen et al. \cite{Nissen92}, Nissen et al. \cite{Nissen01})
based on the forbidden line at $\lambda$ 630.03\,nm suggest, on the other hand,
a more or less flat trend of $\OFe$ in the metallicity 
range of $-2.0 < \feh < -1.0$.
Concerning
other well-studied $\alpha$-capture elements, e.g. Mg, Si
and Ca, earlier observed trends of $\afe$ versus $\feh$
(e.g. Ryan et al. \cite{Ryan96})
seemed to be in  good agreement with a plateau of
$\afe$ $\sim$ 0.4 dex. In particular, Fuhrmann (\cite{Fuhrmann98})
found a clearly flat $\MgFe$ ratio for $\feh < -0.8$
based on a LTE abundance analysis of Mg and Fe.
However, the recent work of Stephens \& Boesgaard (\cite{Stephens02})
based on Keck spectra of more than 50 halo and disk stars suggests
a quasi-linear trend of increasing $\afe$ with decreasing $\feh$
(see their Fig. 20), but we note that their average $\afe$ is only about 
$\sim$ 0.4 dex at $\feh = -3.0$.
In addition, Idiart \& Thevenin (\cite{Idiart00})
have found a complicated structure in
$\MgFe$ and $\CaFe$ vs $\feh$ based on a non-LTE re-analysis of 
equivalent widths from various sources.
Although the scatter is significant for $\feh < -1.0$, 
their results are consistent with plateau-like mean trends 
at $\sim +0.2-0.3$ for both $\MgFe$ and $\CaFe$
for halo stars.
Since different trends of $\afe$ vs. $\feh$ correspond
to different nucleosynthesis histories of the $\alpha$ elements,
a clarification of the divergency on the 
trend of $\OFe$ vs. $\feh$ is very important for
many astrophysical fields including stellar evolution theory,
nucleosynthesis theory,
the formation and chemical evolution of the Galaxy, and the
age of the Galaxy and the Universe.

Another $\alpha$ element, sulphur, is highly interesting
as an independent tracer of the nucleosynthesis of $\alpha$ elements
and a well-established trend of $\SFe$ vs. $\feh$ could be
helpful in clarifing the discrepancy of the $\OFe$ ratio.
Furthermore, sulphur is of particular interest in 
connection with studies of the chemical enrichment of
damped Ly$\alpha$ systems being one of the few elements
that is not depleted onto dust (Centuri\'{o}n et al. \cite{Centurion00}).
Unfortunately, sulphur has been ignored for many years probably
due to the lack of measurable lines in the optical spectra.
Presently, our knowledge of sulphur abundances 
is limited to only a few studies: Clegg et al. (\cite{Clegg81}), Fran\c{c}ois
(\cite{Francois87}, \cite{Francois88}) and recent works by
Israelian \& Rebolo (\cite{Israelian01}) and 
Takada-Hidai et al. (\cite{Takada02}). Again, the disagreement 
on the $\SFe$ ratio at low metallicities appears:
Fran\c{c}ois (\cite{Francois87}, \cite{Francois88}) suggested 
the $\SFe$ ratio to be constant for stars with $-1.6 < \feh <-1.0$, while Israelian \& Rebolo
(\cite{Israelian01}) favor a linear rise of $\SFe$ ratio 
with decreasing $\feh$ in the metallicity range of $-3.0 < \feh <-0.6$. 
With a smaller slope ($-$0.25 vs $-$0.46 in Israelian \& Rebolo paper) the work of Takada-Hidai et al. (\cite{Takada02}) support a
linear trend of $\SFe$.

Sulphur and iron abundances in disk stars are usually better
determined than in the case of halo stars.
The sulphur lines have a suitable strength for
accurate abundance determination 
and $\Fetwo$ lines, which have very small non-LTE effects,
can be used for deriving the iron abundance.
Therefore, the $\SFe$ vs. $\feh$
trend can be more reliably established. With this advantage,
the determination of sulphur abundances in disk stars with $\feh > -1.0$ is
important especially from the nucleosynthesis theory
point of view. The comparison of the sulphur
behaviour with those of other $\alpha$ elements in disk stars
provides a key clue to the genuine origin of sulphur and
further to the chemical evolution of the Galaxy.
As the nearest even-$Z$ neighbour of sulphur in the
periodic table of elements, silicon is very suitable as
a representative of the $\alpha$ elements for
such a comparison. 

In the present work, we investigate the
behaviour of sulphur in disk stars and the nucleosynthesis
origin of this element. The investigation takes benefit of more accurate
observations of sulphur lines, better determinations 
of stellar parameters, improved atmospheric models, and hence   
more reliable derivations of sulphur, silicon and iron 
abundances than in previous works (Clegg et al. \cite{Clegg81}; Fran\c{c}ois
\cite{Francois87}, \cite{Francois88}), where the scatter of
$\SFe$ for disk stars is so large that it is not clear if
sulphur really behaves as an $\alpha$-capture elements or not.
In a following paper (Nissen et al. \cite{Nissen02}), the study
will be extented to halo stars such that the trend
of $\SFe$ will be covered for the whole metallicity range
of  $-3.0 < \feh < +0.2$.

\section{Observations}
The stars were selected from the $uvby\beta$ photometric catalogues
of Olsen (\cite{Olsen83}, \cite{Olsen93}) according to the
criteria of $5600 \leq \teff \leq 6600~\mathrm{K}$, 
$3.8 \leq \logg \leq 4.5$ and $ -1.0 \leq \feh \leq +0.5$
with approximately equal numbers of stars in every metallicity
interval of 0.1 dex. This selection aimed at obtaining a
sample of stars with similar parameters in order to minimize
the analysis uncertainties including the ways of determining
stellar parameters and atmospheric modeling.

\begin{figure}
\resizebox{\hsize}{!}{\includegraphics{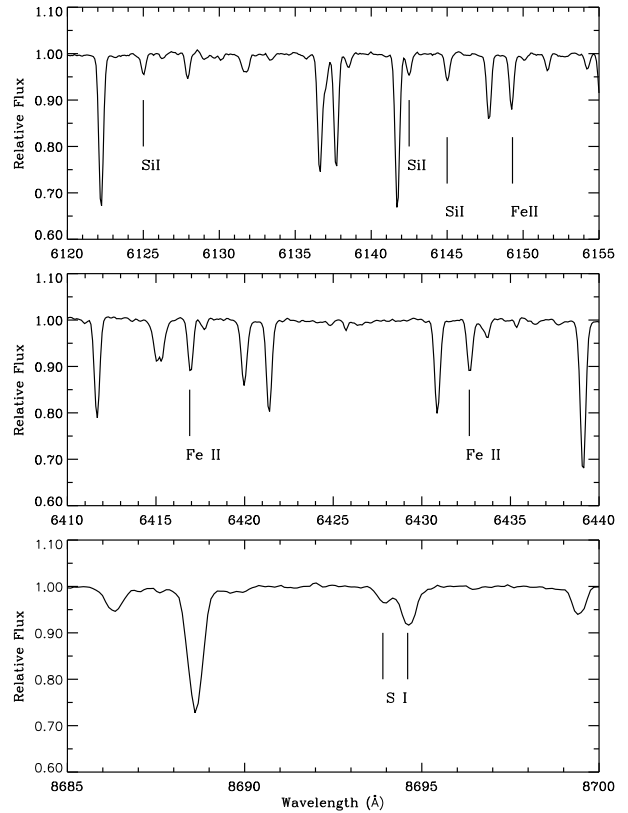}}
\caption{Three sections of the spectrum for HD\,168151
($\teff=6530$~K, $\logg=4.12$, $\feh=-0.28$) as observed with
the 2.16m telescope at Xinglong Station.} 
\label{fig:sp}
\end{figure}

\begin{figure}
\resizebox{\hsize}{!}{\includegraphics{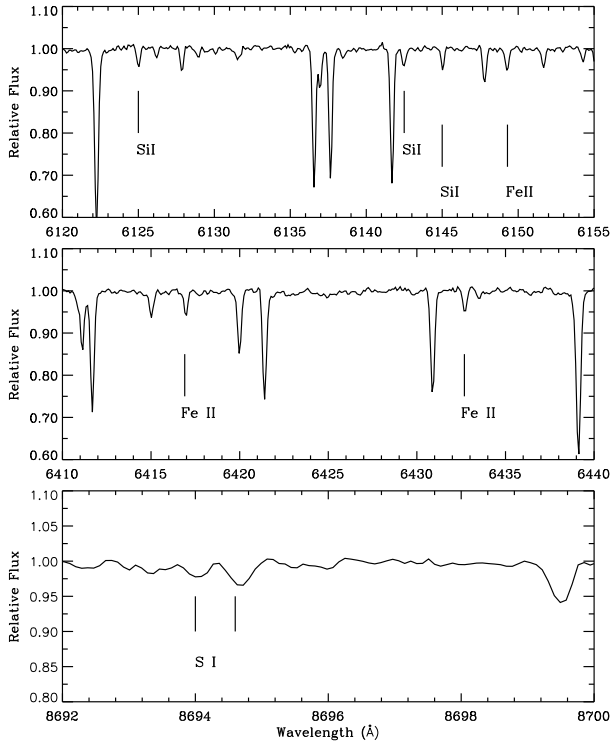}}
\caption{Three sections of the spectrum for HD\,22879 
($\teff=5788$~K, $\logg=4.29$, $\feh=-0.91$) as observed with
the 2.16m telescope at Xinglong Station.}
\label{fig:sp2}
\end{figure}

Spectra of stars were obtained during three observing
runs: Oct. 17-22, 2000, Jan 8-10, 2001 and Aug. 11, 2001 with
the Coud\'{e} Echelle
Spectrograph attached to the 2.16m telescope at National
Astronomical Observatories of Chinese Academy of Sciences
(Xinglong, China). The detector was a Tek CCD 
($ 1024\times 1024$ pixels with $24\times 24~\mu m^{2}$ each in
size). The red arm of the spectrograph with a 31.6 grooves/mm grating
was used in combination with a prism as cross-disperser, providing a
good separation between the echelle orders.  With a 0.5 mm slit
(1.1 arcsec), the resolving power was on the order of 37\,000.
The exposure time was chosen in order to obtain a
signal-to-noise around 150-250 and the spectral coverage 
is 580--880\,nm except for 3 stars with the wavelength
range 550--830\,nm. As examples, three sections of the
spectrum for HD\,168151 and HD\,22879
are shown in Fig.~\ref{fig:sp} and Fig.~\ref{fig:sp2}
with $\Fetwo$,
$\Sione$ and $\Sone$ lines marked by vertical lines.
Note that HD\,22879 is the most metal-poor star
in this sample, and the weak $\Sone\,869.39$ line was
not used in deriving the sulphur abundance because
the noise level in the continuum at this region
is comparable with the line strength.

The data reduction, using MIDAS software, 
involved standard routines for order definition,
background subtraction, wavelength calibration, flat-field division,
spectrum extraction followed by radial velocity shift correction
and continuum normalization. The equivalent widths (EWs) were 
measured by direct integration or Gaussian fitting depending
on which method gave the more reliable value. 
The two
$\Sone$ lines at $\lambda869.4$\,nm were separately measured 
using a combination of two Gaussians. 

Spectroscopic binaries and
a few stars with a high-rotation of $V_{\rm rot} \sin i \geq 25 \kmprs $ 
were excluded from the sample leaving finally 26 stars for
abundance analysis. Table~\ref{tb:all} presents the star sample
as well as the stellar parameters, kinematics, the
equivalent widths of the $\Sone$ lines and abundances.

\section{Analysis}
\subsection{Stellar parameters}
The effective temperature was estimated from Str\"omgren photometric
indices (Olsen \cite{Olsen83}, \cite{Olsen93}) 
by using the $b-y$ calibration of 
Alonso et al. (\cite{Alonso96}).
We have calculated the $E(b-y)$ for every program star 
using the calibrations of Olsen (\cite{Olsen88}) but 
found no significant reddening correction.
All stars have $E(b-y)$ less than 0.010.

$K$ magnitudes for 4 stars are available in the $2MASS$ survey
(Finlator et al. \cite{Finlator00}).
We derived the temperature by using the $V-K$ calibrations of 
Alonso et al. (\cite{Alonso96}). The results are
5698~K for HD\,22879,
6077~K for HD\,28620, 6232~K for HD\,77967 and 5882~K for HD\,216106.
These values are
slightly lower than those derived from the $b-y$ indices
(see Table~\ref{tb:all})
with a mean deviation $<T(V-K)-T(b-y)>= -66\pm$27 K, but
they are consistent within the estimated error of the temperature 
determination, i.e. $\sim$100~K.
For consistency, we adopted the value from the $b-y$ indices for
all stars. Furthermore, we checked that there is no
significant trend of 
the derived iron abundance with excitation potential of low 
energy level of lines, supporting the adopted temperatures
to within $\pm100$~K.

\begin{table*}
\caption{Stellar basic parameters, kinematics, equivalent widths
of the $\Sone$ lines and abundances}
\label{tb:all}
\setlength{\tabcolsep}{0.06cm}
\begin{tabular}{lrrrrrrrrrrrrrr}
\hline
\noalign{\smallskip}
  HD  & $\teff$ & $\logg$ & $\xi_t$ & $\Ulsr$ & $\Vlsr$ & $\Wlsr$ &
EW(1) & EW(2) &EW(3) & EW(4) & EW(5)& $\ffe$ & $\sfe$ & $\sife$ \\[-1.8mm]
 & &  &\multicolumn{4}{c}{------------------------} & & & \\[-2.0mm]
   & K &   &  & \multicolumn{3}{c}{$\kmprs$} &m\AA &m\AA &m\AA &m\AA &m\AA & & & \\
\hline
\noalign{\smallskip}
   693 &6173&4.11& 1.6&$-$29& $-$8&$-$12&   9.8&   --&   -- & 10.7& 30.1&$-$0.30& $-$0.06& $-$0.04\\
  1461 &5660&4.33& 0.6& 22&$-$34&  5&   -- & 16.9&   -- & 13.4& 35.9& 0.47& $-$0.20& $-$0.21\\
  9826 &6119&4.12& 1.9&$-$38&$-$17& $-$8&  -- &  -- &   -- & 27.8& 56.2& 0.12&  0.04& $-$0.03\\
 10453 &6368&3.96& 2.1&$-$46&$-$59& $-$3&  12.3& 11.1&  19.2& 16.2& 52.7&$-$0.46&  0.17&  0.15\\
 13540 &6301&4.12& 1.7& $-$2& 29&  5& 13.1& 20.6&  -- & -- &   -- & $-$0.43&  0.19&  0.08\\
 16895 &6228&4.27& 1.6& 21&  7&  7&  19.4& 16.2&   -- &  -- &51.0 & 0.01&  0.01& $-$0.04\\
 17948 &6455&4.20& 1.6& 21& 14& 20&  13.8& 11.9&  22.6&  -- &43.9 &$-$0.26&  0.03&  0.02\\
 22879 &5788&4.29& 1.4&100&$-$81&$-$36&   -- &  -- &   -- &  -- &10.5 &$-$0.91&  0.23&  0.23\\
 28620 &6114&4.11& 1.6& 11&  5&  8&  9.9&  9.3&  13.2& 9.6 &33.4 & $-$0.38&  0.06& $-$0.01\\
 30652 &6424&4.28& 1.8& 15& $-$10& 12& 23.8& 17.6&   -- &  -- & -- &  0.04& $-$0.02& $-$0.04\\
 33564 &6276&4.16& 1.9&$-$30& 11&  3& 24.9& 20.7&   -- & 27.5& -- & 0.05&  0.05&  0.04\\
 39315 &6202&3.83& 1.8&$-$36& 33&  6&  -- & 10.7&  22.9& 17.4& 45.1& $-$0.38&  0.16&  0.10\\
 49933 &6592&4.21& 1.9&$-$34& $-$9& $-$3&  -- & 12.0&  18.3&  -- & 40.5&$-$0.42&  0.10&  0.05\\
 58461 &6562&4.11& 2.0&  8&  9&$-$22&   -- & 14.6&   -- &  -- & 49.8&$-$0.15& $-$0.05& $-$0.08\\
 77967 &6329&4.15& 2.0&$-$74&  4& 11& 12.1&  8.5&  15.3&  -- & 35.4& $-$0.46&  0.11&  0.11\\
 82328 &6308&3.84& 1.7& 47&$-$30&$-$17& 16.4& 13.3&   -- & 18.8& 45.5& $-$0.13& $-$0.09& $-$0.04\\
 84737 &5813&4.12& 1.5&$-$22& 0& 24&   -- &  -- &   -- & 14.2& 38.2& 0.13& $-$0.10&  0.00\\
 94388 &6379&3.96& 1.8&$-$39&$-$12&$-$10&  27.7& 25.1&   -- &  -- & 71.7& 0.07&  0.05&  0.03\\
150177 &6061&3.93& 1.7& $-$4&$-$19&$-$18&  -- &  7.4&  14.4&  -- &  -- & $-$0.63&  0.17&  0.10\\
159307 &6237&3.93& 1.7&  3&$-$18& 10&  10.4&  8.1&  14.3&  -- &  -- &$-$0.65&  0.24&  0.20\\
162003 &6498&4.02& 2.1&$-$40&  1&  1&  22.6&  -- &   -- &  -- & 61.2& 0.02& $-$0.08& $-$0.04\\
168151 &6530&4.12& 1.9& $-$4& $-$9&$-$44& -- &  -- &  26.7& 21.3& 44.5& $-$0.28&  0.09&  0.05\\
187013 &6298&4.15& 1.9&$-$47& $-$2&$-$17& 18.4&  -- &  26.3& 21.6& 41.4& $-$0.02& $-$0.09& $-$0.04\\
215648 &6158&3.96& 1.6&$-$14&$-$27&$-$21& 13.1& 11.2&  23.1&  -- &  42.8& $-$0.24&  0.05&  0.04\\
216106 &5923&3.74& 1.5&$-$47&$-$34&$-$25&   -- &  -- &  21.6&  -- &  38.2&$-$0.16&  0.04&  0.03\\
222368 &6178&4.08& 1.7& $-$2&$-$21&$-$20& 18.2& 14.7&   -- &  -- &  44.9& $-$0.13&  0.06&  0.03\\
\hline
\end{tabular}
\begin{list}{}{}
\item[]EW: (1)$\lambda$604.603\,nm; (2)$\lambda$605.267\,nm; 
(3)$\lambda$675.717\,nm; 
(4)$\lambda$869.396\,nm; (5)$\lambda$869.464\,nm 
\end{list}
\end{table*}

As the gravity determines the degree of ionization 
for each element and it has a significant effect
on $\Fetwo$ and $\Sone$ lines, a careful derivation of this parameter
is important.
With the available Hipparcos parallaxes,  we can determine 
accurate gravities from the relations $ g \propto {\mathcal M}/R^{2} $ 
and $ L \propto R^{2}\teff^{4}$. The stars have accurate
parallaxes with a relative error of $< 0.05$ except for
3 stars with $\frac{\Delta \pi}{\pi} \sim 0.08$.
Using the derived temperature and absolute magnitude, the mass was
estimated from the stellar position in the $ M_{V} - \logteff$ diagram 
by interpolating in the evolutionary tracks of VandenBerg 
et al. (\cite{VandenBerg00}) and the bolometric correction 
was interpolated from the grids of Alonso et al. (\cite{Alonso95}).

The metallicity, required in the input for temperature 
determination and abundance calculation, was first derived from 
the Str\"omgren $m_1$ index using the calibrations
of Schuster \& Nissen (\cite{Schuster89}).
But the final metallicity from the abundance analysis 
of the $\Fetwo$ lines
was used to iterate the whole procedure to full consistency. 
The microturbulence, $\xi_t$, was
obtained by requiring a zero slope of $\feh$ vs. EW based on 
$\Fetwo$ lines.

The uncertainties of the stellar parameters are 
estimated to be around 100 K in $\teff$, 0.1 dex in $\logg$, 0.1 dex
in \feh\ and 0.3\kmprs\ in $\xi_t$.

\subsection{Abundance calculations}

The abundance analysis is based on a grid of flux constant, 
homogeneous, LTE model atmospheres , which
were computed with a recent version of
the MARCS code using updated continuous opacities 
including UV line blanketing of millions of absorption lines
and many molecular lines (Asplund et
al. \cite{Asplund97}). These models are based on the assumption
of $\afe=0.0$. Differences are only 0.05 dex in $\feh$,
$-$0.04 dex in $\SiFe$ and $-0.02$ in $\SFe$ by using a model calculated
with $\afe=0.4$ for HD\, 22879, which
is the most metal-poor star in our sample.
Given that enhancements of $\SFe$ and $\SiFe$ for our sample of stars
are less than 0.3 dex, the enhancement of $\afe$ or not
in the model atmosphere has a negligible effect on the result.
The abundance analysis program, EQWIDTH,
was used to calculate the theoretical equivalent widths 
from the models, and elemental
abundances were derived by requiring that the calculated
equivalent widths  should match the observed ones.
When more than one line of the same species were
measured for a star, the mean value is adopted by
giving equal weight to each line.

Log$gf$ values for the $\Sone$, $\Sione$ and $\Fetwo$
lines are presented in Table~\ref{tb:line} together with
solar equivalent widths
measured from the Moon spectrum observed at
Xinglong except for three lines 
with $\lambda > 860$\,nm.
The equivalent widths
for these three lines were measured from the solar
flux spectrum of Kurucz et al. (\cite{Kurucz84}) because
the Moon spectrum observed at Xinglong does not 
cover the wavelength range of $\lambda > 820$\,nm.
Solar abundances are calculated based on a
solar model atmosphere computed with the
same code as used for the stars. 
The $gf$-values from Bi\'emont 
et al. (\cite{Biemont91}) (including the 
normalization of $-$0.05 dex to get agreement
with lifetime measurements)
are adopted for most $\Fetwo$ lines.
With these values, we derived
a solar Fe abundance close to the meteorite value
$\nfes=7.50$ (Grevesse \& Sauval \cite{Grevesse98}).
We obtained differential
log$gf$ values for the remaining three $\Fetwo$ lines by forcing them to
give a consistent solar abundance.
The log$gf$ values for
four of the five $\Sone$ lines are taken from Kurucz (\cite{Kurucz02}).
As shown in Table~\ref{tb:line}, the line-to-line 
scatter of the derived sulphur abundance is small 
for these four lines. From the $\lambda$604.60\,nm line, we
get a higher solar sulphur abundance.
In the solar flux spectrum, this line seems to be
blended with a weak nearby line at the blue edge, but
we are not sure if this can account for the
divergent abundance because 
the $\lambda$604.60\,nm line seems to give a consistent abundance with that
derived from the other $\Sone$ lines for most of our sample
stars. The log$gf$ value presented in Kurucz (\cite{Kurucz02})
($-0.78$) will give an even higher abundance than that
presented in Table~\ref{tb:line}. 
Therefore, we derived a differential log$gf$ value for 
this line based on the spectrum of HD\,693, which 
has a high signal-to-noise above 300.
We get $\nss =7.20$ (excluding the $\lambda$604.60\,nm line), which
is the same as the meteorite
value (Grevesse \& Sauval \cite{Grevesse98}).
We emphasize that the inclusion of this line or not
in the abundance analysis for stars will not change the
result of this work because it gives a consistent abundance with that
derived from other lines.
Experimental log$gf$ values for five
$\Sione$ lines are taken from Garz (\cite{Garz73}), and
the differential log$gf$ values for the remaining
$\Sione$ lines are taken from Chen et al. (\cite{Chen00}).
The enhancement factor, $E_{\gamma}$, is assumed to be 2.5 
for $\Fetwo$ and $\Sone$ and 1.3 for $\Sione$ lines.
There is no significant difference in the derived sulphur and iron abundances
if $E_{\gamma}$ is changed from 2.5 to 1.5, because the
$\Sone$ and $\Fetwo$ lines are weak. $\Sione$ lines 
are somewhat stronger, but the maximum effect on
$\SiFe$ for the most metal-rich star HD\,1461 is only 0.05 dex
when $E_{\gamma}$=2.5 is adopted instead of 1.3.

In sum, we obtained mean solar values:
$\feh =0.02$, $\SFe=-0.02$ and $\SiFe=-0.01$
relative to the meteorite abundances given by
Grevesse \& Sauval (\cite{Grevesse98}). Stellar
abundances, presented in Table~\ref{tb:all},
are relative to these values, so that the solar ratios $\XFe$
in the following figures are exactly zero.

\begin{table}
\caption{The atomic line data for $\Sone$, $\Sione$ and $\Fetwo$ lines,
and solar flux equivalent widths and abundances. Note that 
the mean solar sulphur abundance
is derived excluding the $\lambda$604.60\,nm line.}
\label{tb:line}
\begin{tabular}{lrrrrrrr}
\hline
\noalign{\smallskip}
Ele &$\lambda$ & $\chi_{l}$ &$\log gf$ & ref.& $E_{\gamma}$ & EW & Abu.\\
    & nm    & eV &  &    &  &  m\AA &   \\
\hline
$\Sone$ & 604.603&7.87& $-$0.51&1&2.5& 14.5& 7.32: \\
$\Sone$ & 605.267&7.87& $-$0.63&2&2.5&  9.8 & 7.25 \\
$\Sone$ & 675.717&7.87& $-$0.31&2&2.5& 16.1& 7.21\\
$\Sone$ & 869.396&7.87& $-$0.52&2&2.5& 11.0& 7.19\\
$\Sone$ & 869.464&7.87&  0.05&2&2.5& 28.9& 7.17\\
\multicolumn{8}{c}{Mean $\nss=7.20\pm0.03$} \\
\noalign{\smallskip}
$\Sione$& 566.556&4.92& $-$2.04&3&1.3& 40.9& 7.57\\
$\Sione$& 569.043&4.93& $-$1.87&3&1.3& 53.8& 7.62\\
$\Sione$& 570.111&4.93& $-$2.05&3&1.3& 37.1& 7.52\\
$\Sione$& 570.840&4.95& $-$1.40&1&1.3& 76.0& 7.49\\
$\Sione$& 577.215&5.08& $-$1.66&1&1.3& 54.8& 7.55\\
$\Sione$& 579.308&4.93& $-$1.95&1&1.3& 47.2& 7.59\\
$\Sione$& 579.786&4.95& $-$2.05&3&1.3& 43.1& 7.64\\
$\Sione$& 594.855&5.08& $-$1.19&1&1.3& 87.1& 7.57\\
$\Sione$& 612.503&5.61& $-$1.54&1&1.3& 33.3& 7.54\\
$\Sione$& 614.249&5.62& $-$1.48&1&1.3& 38.7& 7.59\\
$\Sione$& 614.502&5.62& $-$1.43&1&1.3& 39.6& 7.55\\
$\Sione$& 703.491&5.87& $-$0.81&1&1.3& 69.6& 7.57\\
$\Sione$& 722.621&5.61& $-$1.30&1&1.3& 47.6& 7.53\\
$\Sione$& 740.579&5.61& $-$0.68&1&1.3& 93.6& 7.54\\
$\Sione$& 741.596&5.61& $-$0.71&3&1.3& 91.5& 7.54\\
$\Sione$& 791.838&5.95& $-$0.54&1&1.3& 95.3& 7.65\\
$\Sione$& 793.235&5.96& $-$0.35&1&1.3&113.6& 7.64\\
$\Sione$& 872.802&6.18& $-$0.36&1&1.3& 91.2& 7.53\\
\multicolumn{8}{c}{Mean $\nsis$=7.57$\pm0.03$} \\
\noalign{\smallskip}
$\Fetwo$& 599.138&3.15& $-$3.61&4&2.5& 31.9& 7.52\\
$\Fetwo$& 608.410&3.20& $-$3.86&4&2.5& 20.7& 7.52\\
$\Fetwo$& 611.332&3.22& $-$4.16&1&2.5& 11.3& 7.50\\
$\Fetwo$& 614.925&3.89& $-$2.77&4&2.5& 37.5& 7.50\\
$\Fetwo$& 617.940&5.57& $-$2.65&4&2.5&  3.2& 7.50\\
$\Fetwo$& 623.995&3.89& $-$3.49&4&2.5& 13.3& 7.53\\
$\Fetwo$& 624.756&3.89& $-$2.38&4&2.5& 56.4& 7.52\\
$\Fetwo$& 638.371&5.55& $-$2.32&4&2.5&  7.2& 7.54\\
$\Fetwo$& 641.693&3.89& $-$2.79&4&2.5& 40.7& 7.59\\
$\Fetwo$& 643.268&2.89& $-$3.63&5&2.5& 41.4& 7.49\\
$\Fetwo$& 645.639&3.90& $-$2.13&4&2.5& 65.2& 7.46\\
$\Fetwo$& 651.608&2.89& $-$3.38&1&2.5& 54.0& 7.50\\
$\Fetwo$& 722.240&3.89& $-$3.35&4&2.5& 18.9& 7.57\\
$\Fetwo$& 722.446&3.89& $-$3.29&4&2.5& 19.7& 7.54\\
$\Fetwo$& 744.933&3.89& $-$3.22&4&2.5& 21.1& 7.51\\
$\Fetwo$& 747.970&3.89& $-$3.64&4&2.5&  9.7& 7.49\\
$\Fetwo$& 751.583&3.90& $-$3.48&4&2.5& 13.1& 7.50\\
$\Fetwo$& 771.173&3.90& $-$2.59&4&2.5& 49.3& 7.54\\
\multicolumn{8}{c}{Mean $\nfes$=7.52$\pm0.04$} \\
\noalign{\smallskip}
\hline
\end{tabular}
\begin{list}{}{}
\item[] References for $\log gf$ values: (1) differential values in this work;
(2) Kurucz \cite{Kurucz02};
(3) Garz \cite{Garz73};
(4) Bi\'emont et al. \cite{Biemont91};
(5) Hannaford et al. \cite{Hannaford92}
\end{list}
\end{table}

\begin{table}
\caption{Dependences of relative abundances on the stellar parameters.}
\label{tb:abuerr}
\begin{tabular}{lrrrr}
\hline
\multicolumn{5}{c}{HD28620} \\
\multicolumn{5}{c}{$\teff$=6114~K,  $\logg$=4.11,  $\feh$=$-$0.38, $\xi_t$=1.6 $\kmprs$} \\
\hline
 & $\teff$ & $\logg$ &$\feh$ & $\xi_t$  \\
   &+100~K & +0.1 dex  &+0.1 dex& +0.3$\kmprs$ \\
\hline
\noalign{\smallskip}
$[\Feone/{\rm H}]$   & 0.07 &0.00 &0.00 &$-$0.05   \\
$[\Fetwo/{\rm H}]$   &$-$0.01 & 0.04 & 0.01 &$-$0.04  \\
$[\Sone/\Feone]$   &$-$0.12 & 0.03 & 0.00 & 0.05  \\
$[\Sone/\Fetwo]$   &$-$0.04 &$-$0.01 &$-$0.01 & 0.04  \\
$[\Sione/\Feone]$  &$-$0.04 & 0.00 & 0.00 & $-$0.03  \\
$[\Sione/\Fetwo]$  & 0.04 &$-$0.04 &$-$0.01 &$-$0.02  \\
\hline 
\end{tabular}
\end{table}

\subsection{Abundance uncertainties}
In order to investigate which lines, 
$\Feone$ or $\Fetwo$, are the best indicator
for the iron abundance in deriving the $\SFe$ ratio,
we show in Table~\ref{tb:abuerr} the sensitivity of the
derived abundances to the stellar parameters
for a representative star \object{HD\,28620}.
This star has $\teff = 6114$~K, $\logg = 4.11$, 
$\feh = -0.38$, which is typical for the range of
temperature, gravity, and metallicity
of our sample of stars. In these calculations, the
log$gf$ values for $\Feone$ lines are the same
as those of Chen et al. (\cite{Chen00}), and
the deviations are obtained by a change of
100~K in effective temperature, 0.1 dex in gravity, 
0.1 dex in metallicity, and 0.3 $\kmprs$ in 
microturbulence.

Clearly, sulphur abundances with respect to
iron abundances derived from $\Fetwo$ lines are more
immune to variations of atmospheric parameters 
as compared to sulphur relative to iron
from $\Feone$ lines.
This is primarily due to the high ionization 
potential of sulphur (10.36\,eV), which means that
sulphur mainly occurs as $\Sone$ atoms at
temperatures of $\sim$6000~K.
Thus, we adopt the iron abundance derived from
$\Fetwo$ lines (hereafter the iron abundance)
for all stars. Since the
total error of $\SiFe$ ratio is only a bit larger
when Si is measured relative to Fe derived from
$\Fetwo$ lines instead of $\Feone$ lines
(see Table~\ref{tb:abuerr}),  
both silicon and sulphur abundances
are relative to iron abundances from
$\Fetwo$ lines. A further advantage of using
$\Fetwo$ lines is that non-LTE effects on $\feh$
are bound to be small because nearly all iron
occur as $\Fetwo$ ions --- a conclusion which is supported
by detailed  calculations by Th\'evenin \& Idiart (\cite{Thevenin99}).

The line-to-line scatter of the deduced
abundances 
gives an estimate of the uncertainty arising
from the measured equivalent widths. 
For silicon and iron, more than 10 lines
are available for most stars and the statistical error of the mean abundance
is less than 0.02 dex.
With 2-5 sulphur lines used in this work, the corresponding error 
of the mean abundance is of the order of 0.03 dex.
Taking into account the errors from the
uncertainties of the stellar parameters in Table~\ref{tb:abuerr},
we estimate the total errors to be $\sim$0.06 dex for $\feh$ and $\SiFe$,
and $\sim$0.07 dex for $\SFe$; the
corresponding error bars are shown in Fig.~\ref{fig:abu}.

The above error estimates do not, however, account for possible systematic
errors introduced for example by the assumption of LTE or
the use of 1D hydrostatic model atmospheres.
The non-LTE effect on the $\Sone$ $\lambda$869.39\,nm
and $\lambda$869.46\,nm lines has been investigated
by Takada-Hidai et al. (\cite{Takada02}), who showed that 
this effect is negligible. For the other three $\Sone$
lines, non-LTE effects are also expected to be insignificant because
they have the same excitation potential as the
$\lambda$869.46\,nm line and they are weaker.
As mentioned above, iron abundances derived 
from $\Fetwo$ lines 
are generally immune to non-LTE effects. Thus,
non-LTE effects are expected to be negligible in the trend of $\SFe$ 
versus $\feh$.

Following the same procedure as in Asplund et al. (\cite{Asplund99})
and Asplund \& Garc\'{\i}a P{\'e}rez (\cite{Asplund01}), we have estimated
the effects of stellar granulation on the derived S and Si
abundances using the new generation of ab-initio 3D hydrodynamical
model atmospheres (e.g. Asplund et al. \cite{Asplund00}). 
A strictly differential
comparison of the predicted line strengths in the 3D models and
1D MARCS model atmospheres with identical parameters has been 
performed for S\,{\sc i} and Si\,{\sc i} lines; the results for 
Fe\,{\sc ii} lines have been culled from Asplund \& Garc\'{\i}a P{\'e}rez 
(2001). As seen in Table~\ref{t:3D}, the 1D results are relatively 
little affected due to the high excitation potential of the 
S\,{\sc i} and Si\,{\sc i} lines which forces them to be formed at
similar atmospheric depths as the Fe\,{\sc ii} lines. For the relevant
metallicity range, the S/Fe and Si/Fe ratios differ by $\le 0.04$\,dex
when using 3D model atmospheres compared to the 1D case. 
To avoid interpolation within the still limited sample 
of granulation corrections and given their minor impact on the 
final results, we have made no attempt to apply such corrections 
to the 1D abundances of the individual stars in the following discussion.

\begin{table}[t]
\caption{Comparison of the Si, S and Fe LTE abundances derived with 1D 
hydrostatic and 3D hydrodynamical model atmospheres. 
The 3D abundances 
are those which reproduce the equivalent widths computed using a 
1D model atmosphere with identical parameters, 
a microturbulence of $\xi_{\rm turb} = 1.0$\,km\,s$^{-1}$ 
and the abundances given in the third and fourth columns
}
\label{t:3D}
\begin{tabular}{lccccc} 
 \hline 
\noalign{\smallskip}
$T_{\rm eff}$ & log\,$g$ & [Fe/H] & [$\alpha$/Fe]$^{\rm a}$ & 
$\Delta$ (S/Fe)$^{\rm b}$ &
$\Delta$ (Si/Fe)$^{\rm c}$ \\
 $$[K] & [cgs] & & &  
\smallskip \\
\hline 
\noalign{\smallskip}
5767 & 4.44 & $+0.0$ & $+0.00$ & $-0.02$ & $-0.01$ \\
5822 & 4.44 & $-1.0$ & $+0.40$ & $-0.03$ & $-0.04$ \\
6191 & 4.04 & $+0.0$ & $+0.00$ & $+0.02$ & $-0.01$ \\
6180 & 4.04 & $-1.0$ & $+0.40$ & $+0.01$ & $-0.03$ \\
\hline
\end{tabular}
\begin{list}{}{}
\item[$^{\rm a}$] The $\alpha$-enhancement for the S and Si abundances. 
For these calculations, log\,$\epsilon$(S)=7.33, 
log\,$\epsilon$(Si)=7.55 and log\,$\epsilon$(Fe)=7.50 
have been assumed for the Sun following Grevesse \& Sauval (1998).
\item[$^{\rm b}$] Based on the five S\,{\sc i} lines listed in
Table 2 and the six Fe\,{\sc ii} lines used by 
Asplund \& Garc\'{\i}a P{\'e}rez (2001).
\item[$^{\rm c}$] Based on the four Si\,{\sc i} 570.1, 594.8, 722.6
and 872.8\,nm lines and the six Fe\,{\sc ii} lines used by 
Asplund \& Garc\'{\i}a P{\'e}rez (2001).
\end{list}
\end{table}

\section{Results and discussions}
\subsection{Sulphur abundances vs. metallicity}

\begin{figure}
\resizebox{\hsize}{!}{\includegraphics{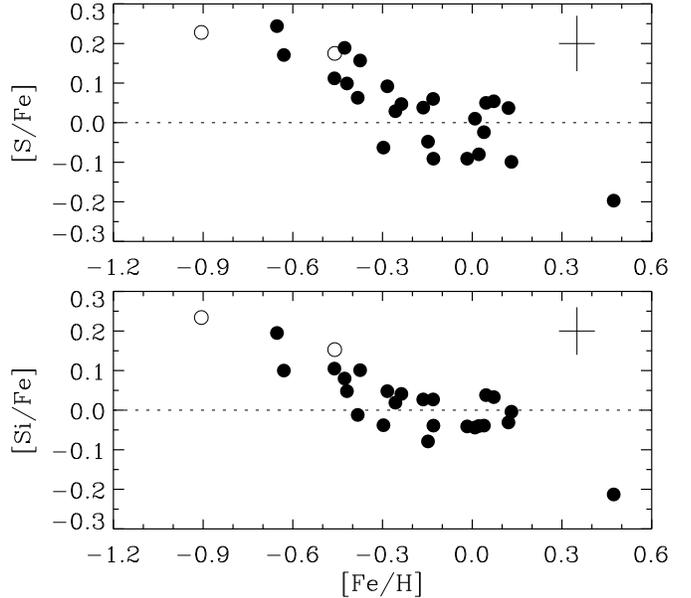}}
\caption{The $\SFe$ and $\SiFe$ ratio as a function of metallicity.
Filled circles denote the thin disk stars while open circles
represent stars with kinematics typical for the thick disk.}
\label{fig:abu}
\end{figure}

The $\SFe$ ratio as a function of metallicity
is shown in Fig~\ref{fig:abu}. It seems that the $\SFe$ ratio 
is nearly solar for $\feh > -0.3$, below which it
increases with decreasing metallicity and an overabundance of $\sim$0.25
dex is indicated at $\feh = -1.0$.
There may be a tendency for decreasing
$\SFe$ for super metal-rich stars but this is
based on one star with $\feh \sim +0.5$ only.
These results are in agreement with those of
Fran\c{c}ois (\cite{Francois87}, \cite{Francois88}),
but the scatter in the $\SFe$ ratio
at a given metallicity is significantly smaller in the present work.

\subsection{Reanalysis of data from Fran\c{c}ois (\cite{Francois87}, \cite{Francois88})}

Israelian \& Rebolo (\cite{Israelian01}) made a
reanalysis of the data presented in
Fran\c{c}ois (\cite{Francois87}, \cite{Francois88})
by using a new determination of stellar parameters and
suggest a linear increase of the $\SFe$ ratio
with decreasing metallicity. One point in their paper
is, however, strange: stars with $\feh > -0.4$ have
$\SFe$ around $-$0.3, which is significantly lower than
the solar ratio. At the same time, they derived a solar
sulphur abundance of 7.20, which is the same
as the meteoritic value. 
Furthermore, the scatter in $\SFe$ obtained by 
Israelian  \& Rebolo (\cite{Israelian01}) 
is significantly larger than in the present work.

\begin{table}
\caption{Stellar parameters and relative abundances for
the reanalysis of data from Fran\c{c}ois (\cite{Francois87}, \cite{Francois88}).}
\label{tb:Slitabu}
\begin{tabular}{rcccccc}
\hline
\noalign{\smallskip}
  HD  & $\teff$ & $\logg$ & $\xi_t$ & $\feh$ & $\feh$ & $\sh$ \\
      &   K     &         &$\kmprs$  & LTE   & NLTE   & LTE    \\
\hline
\noalign{\smallskip}
 59984 &5896&3.93& 1.6& $-$0.88& $-$0.74& $-$0.67\\
 63077 &5825&4.15& 1.2& $-$0.89& $-$0.75& $-$0.57\\
 69897 &6227&4.20& 1.6& $-$0.50& $-$0.41& $-$0.41\\
 76932 &5870&4.08& 1.3& $-$0.98& $-$0.83& $-$0.63\\
 88218 &5661&3.94& 1.4& $-$0.53& $-$0.44& $-$0.26\\
 91324 &6123&3.95& 1.8& $-$0.60& $-$0.50& $-$0.49\\
102365 &5562&4.39& 1.0& $-$0.39& $-$0.32& $-$0.35\\
106516 &6135&4.30& 1.5& $-$0.79& $-$0.66& $-$0.55\\
136352 &5584&4.27& 1.0& $-$0.58& $-$0.48& $-$0.44\\
139211 &6231&4.12& 1.7& $-$0.26& $-$0.21& $-$0.16\\
148816 &5784&4.07& 1.4& $-$1.03& $-$0.87& $-$0.54\\
157089 &5712&4.00& 1.4& $-$0.79& $-$0.66& $-$0.44\\
193901 &5662&4.48& 1.0& $-$1.13& $-$0.96& $-$0.67\\
201891 &5825&4.21& 1.3& $-$1.21& $-$1.03& $-$0.86\\
203608 &6094&4.29& 1.4& $-$0.91& $-$0.77& $-$0.69\\
\hline
\noalign{\smallskip}
\end{tabular}
\end{table}

\begin{figure}
\resizebox{\hsize}{!}{\includegraphics{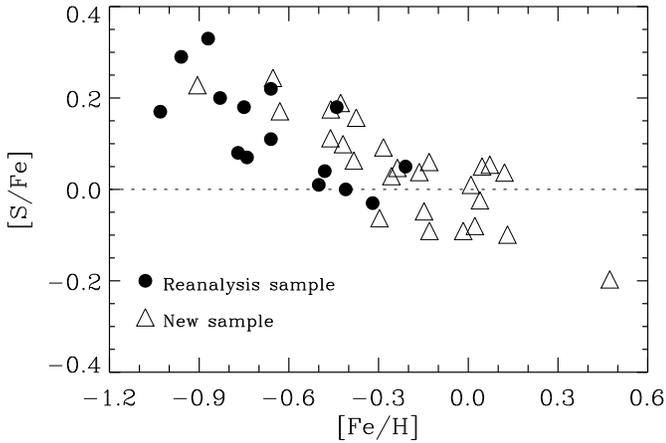}}
\caption{The $\SFe$ vs. $\feh$ trend 
for the new measurements (open triangles) and our
reanalysis of the  Fran\c{c}ois (\cite{Francois87}, \cite{Francois88})
data (filled circles).}
\label{fig:Slitabu}
\end{figure}

\begin{figure}
\resizebox{\hsize}{!}{\includegraphics{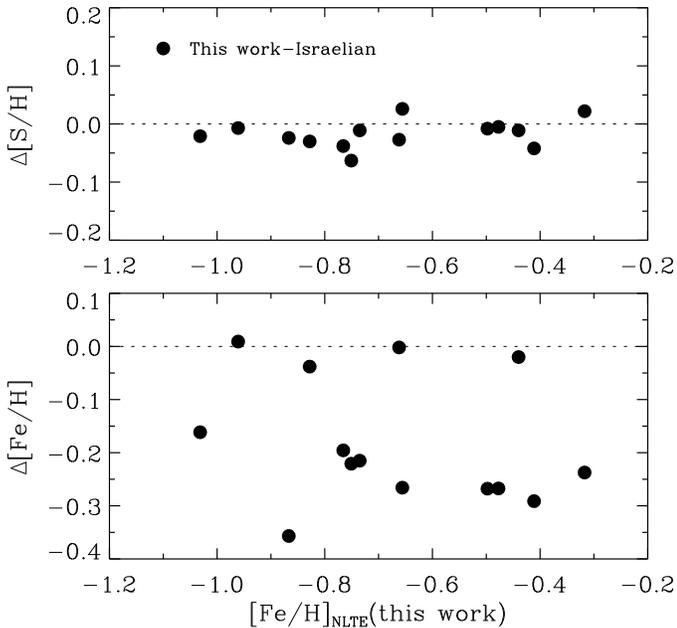}}
\caption{Abundance comparison between this work and
Israelian  \& Rebolo (\cite{Israelian01}) for $\sh$ and $\feh$
based on the reanalysis of the
Fran\c{c}ois (\cite{Francois87}, \cite{Francois88}) stars.}
\label{fig:Slicom}
\end{figure}

In order to investigate possible systematic differences 
relative to the work by Israelian  \& Rebolo (\cite{Israelian01}),
we have made a reanalysis of sulphur and
iron abundances for 15 Fran\c{c}ois (\cite{Francois87}, \cite{Francois88})
stars with similar stellar parameters
as our sample stars, i.e.
$5500 <\teff < 6600$~K, $\logg > 3.8$ and $\feh > -1.1$.
The reanalysis was performed in
the same way as that for our sample of stars, with parameters derived
from  Str\"omgren photometry and Hipparcos parallaxes.
The microturbulence was estimated from the relation presented in 
Edvardsson et al. (\cite{Edvardsson93}) and the equivalent
widths of $\Feone$ lines published by 
Fran\c{c}ois (\cite{Francois87}, \cite{Francois88}) were 
used to derive iron abundances.
Adopted log$gf$ values are $-1.68$ for $\Feone$ 867.48\,nm and $-0.38$ for
 $\Feone$ 869.95\,nm taken from Nave et al. (\cite{Nave94}); the very strong
$\Feone$ 868.86\,nm line was discarded because of its sensitivity to 
the uncertain damping constant.
The solar equivalent widths for the $\Feone$ 867.48\,nm and  $\Feone$ 869.95\,nm
lines were measured from the solar flux spectrum by Kurucz et al. (\cite{Kurucz84}).
Again, stellar abundances (see Table~\ref{tb:Slitabu}) 
are relative to the Sun so that
uncertainties from atomic line data cancel out.
The LTE iron abundances derived from $\Feone$ lines
were corrected for non-LTE effects using the
calculations of Th\'evenin \& Idiart (\cite{Thevenin99}).
These non-LTE iron abundances are more
consistent with those derived from $\Fetwo$ lines
for our new sample of stars.
As shown in Fig.~\ref{fig:Slitabu}, the scatter
in the $\SFe$ vs. $\feh$ trend at a given metallicity
is quite small and the results 
from the reanalysis of the Fran\c{c}ois data
and the new measurements in the present work are consistent.

Fig.~\ref{fig:Slicom} shows the deviations of $\sh$ and $\feh$ 
between the present work and Israelian  \& Rebolo (\cite{Israelian01})
for 14 stars in common. One of the 15 stars
in our reanalysis sample was not included in 
Israelian  \& Rebolo (\cite{Israelian01}).
It is clear that there is no significant
deviation for the sulphur abundances between
the two works, but the iron abundances
show large differences. 
The iron abundances adopted by Israelian  \& Rebolo (\cite{Israelian01}) 
for 10 Fran\c{c}ois (\cite{Francois87}, \cite{Francois88}) stars are
significantly higher than the iron abundances in the present work, 
which explains the low $\SFe$ ratio ($-$0.3) they find 
at solar metallicity. 
We note that they adopt $\feh$ values from different
sources in the literature. Therefore, the iron abundances may be inconsistent
with $\teff$ and $\logg$, and consequently also with the derived $\sh$. This
explains both the large scatter and the offset with respect to the Sun
in Fig. 3 of Israelian  \& Rebolo (\cite{Israelian01}),
emphasizing the need for a proper and consistent treatment of the
different species used to derive abundance ratios such as $\SFe$.
The adopted stellar parameters are, otherwise,  
quite similar for the two works.


\subsection{Correlation between sulphur and silicon abundances}
\begin{figure}
\resizebox{\hsize}{!}{\includegraphics{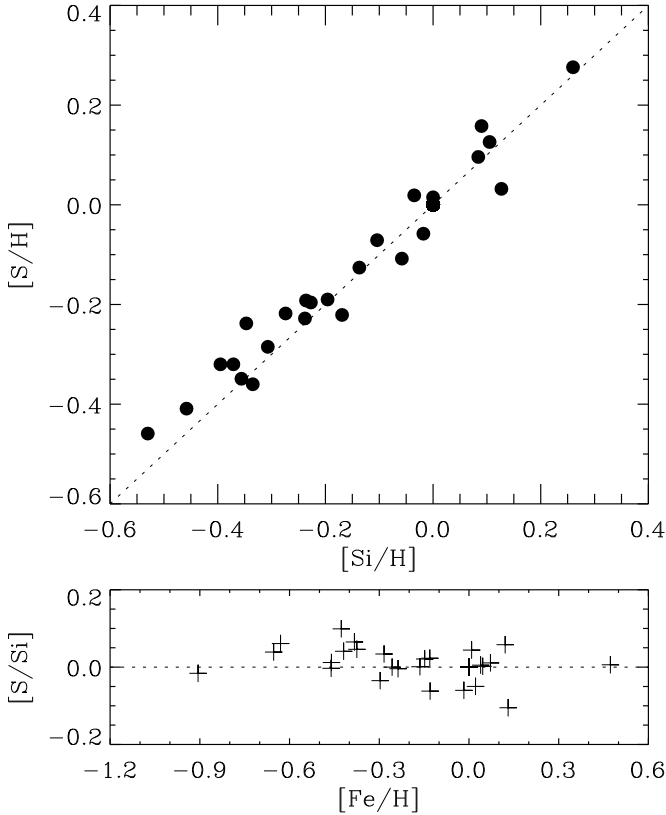}}
\caption{The correlation between $\sh$ and $\sih$, and
the $\ssi$ ratio as a function of metallicity.}
\label{fig:SSi}
\end{figure}

As described in the Introduction, one goal of this
work is the comparison of sulphur with
other $\alpha$-capture elements in particular the closest element, silicon,
being a good representative. The second reason for
the choice of silicon is that we have many good 
silicon lines in the spectra and the oscillator strengths 
are quite reliable; abundances derived from different lines
are consistent for both the Sun and stars.
Thus we can expect that $\SiFe$ vs. $\feh$ will be
well defined.

As shown in Fig.~\ref{fig:abu}, the trend of $\SiFe$ vs. $\feh$ 
is similar to that of sulphur and the scatter at a given
metallicity is also the same.
The one-to-one relation of $\sh$ vs. $\sih$ 
is shown in Fig.~\ref{fig:SSi};
there is a very good correlation between sulphur and
silicon indicating common nucleosynthesis sites
for these two elements. 

In order to investigate the
dependence of the correlation of sulphur and
silicon on metallicity, we plot the $\ssi$ ratio
as a function of $\feh$ in the bottom of Fig.~\ref{fig:SSi}. 
It seems that the
sulphur abundance relative to silicon 
is constant for the whole $\feh$ range investigated
in this work. \\

Goswami \& Prantzos (\cite{Goswami00}) investigated
the chemical evolution of many elements based
on the yields of Woosley \& Weaver (\cite{Woosley95})
for two cases: Case A considered the constant (solar) metallicity yields
while Case B corresponds to
metallicity dependent yields. They found exactly the same trend for
the sulphur and silicon evolution in Case A and Case B in the whole
metallicity range of $-4.0< \feh <+0.5$. 
They suggested that the elements O, Si, S and Ca, behave as true primaries, 
whithout any metallicity dependence of their yields.
The predicted ratios of S and Si in Goswami \& 
Prantzos (\cite{Goswami00}) and other chemical evolution models,
e.g.  Timmes et al. (\cite{Timmes95}), Chiappini et al.
(\cite{Chiappini97}) and Samland et al. (\cite{Samland98}), 
are consistent with the results of the present work for disk stars.

\subsection{Kinematics}

Stellar kinematics are useful information to interpret
the stellar abundances and to probe the origin of
the stars. We calculated
the kinematics for our program stars based on proper
motions and parallaxes in the Hipparcos survey and the radial
velocity derived from the Doppler shift of the spectral lines
in this work.
The calculation of galactic space velocity (U, V, W) follows
that presented in Johnson \& Soderblom (\cite{Johnson87}).
The correction of space velocity to the Local Standard of
Rest is based on a solar motion,
($-$10.0,$+$5.2,$+$7.2) \kmprs\ in (U,V,W) as 
derived from Hipparcos
data by Dehnen \& Binney (\cite{Dehnen98}).

As shown in Table~\ref{tb:all}, most of the sample stars are
thin disk stars ($\Vlsr>-40 \kmprs$). Exceptions are \object{HD\,10453}
and \object{HD\,22879}, which have thick disk kinematics.
The $\SFe$ ratio is about 0.2 dex for both stars, which
is higher than most other stars. These data seem to be in line with 
the result by Edvardsson et al. (\cite{Edvardsson93}), who
found an abundance gradient of $\alpha$ elements
at the metallicity of $\feh \sim-0.7$. But the metallicities
of the two thick disk stars are also the lowest, and 
thin disk kinematics does not imply a low $\SFe$ ratio. 
For example, \object{HD\,77967} and \object{HD\,159307} have
thin disk kinematics, but both $\SFe$ and
$\SiFe$ are close to the enhanced value
of $\sim$0.2 dex. Therefore, it seems that 
these abundances ratios are primarily a function of metallicity 
and the kinematics plays a minor role.

\section{Conclusions}
We have carried out new measurements of sulphur and silicon
abundances for 26 main sequence stars in the metallicity
range of $-1.0 < \feh < +0.5$ and have established
the trends of $\SiFe$ and $\SFe$ with a small scatter at
a given metallicity. The result indicates a very
good correlation between sulphur and silicon
and the correlation does not depend on the metallicity. 
The results provide support for the
chemical evolution models of Goswami \& Prantzos (\cite{Goswami00}), 
Timmes et al. (\cite{Timmes95}),
Chiappini et al. (\cite{Chiappini97}) and 
Samland et al. (\cite{Samland98}) for 
these two elements in the case of disk stars.
It is concluded that the
nucleosynthesis of sulphur is a proxy for silicon,
which is the nearest $\alpha$-capture element
of sulphur in the
periodic table of elements. 

\section*{Acknowledgements}
This research was supported mainly by the Danish Rectors' Conference
and the Chinese Academy of Sciences and partly by the NKBRSF G1999075406
and the NSFC under grant 10173014.


\begin{thebibliography}{}
\bibitem[1995]{Alonso95}
Alonso A., Arribas S., \& Mart\'{\i}nez-Roger C. 1995, A\&A, 297, 197
\bibitem[1996]{Alonso96}
Alonso A., Arribas S., \& Mart\'{\i}nez-Roger C. 1996, A\&A, 313, 873
\bibitem[1997]{Asplund97}
Asplund M., Gustafsson B., Kiselman D., \& Eriksson K. 1997, A\&A, 318, 521
\bibitem[1999]{Asplund99}
Asplund M., Nordlund \AA., Trampedach R., \& Stein R.F. 1999, A\&A, 346, L17
\bibitem[2000]{Asplund00}
Asplund M., Nordlund \AA., Trampedach R., et al. 2000, A\&A, 359, 729
\bibitem[2001]{Asplund01}
Asplund M., \& Garc\'{\i}a P{\'e}rez A.E. 2001, A\&A, 372, 601
\bibitem[1991]{Biemont91}
Bi\'emont E., Baudoux M., Kurucz R.L., et al. 1991, A\&A, 249, 539
\bibitem[1999]{Boesgaard99}
Boesgaard A.M., King J.R., Deliyannis C.P., \& Vogt S.S. 1999, AJ, 117, 492
\bibitem[2000]{Centurion00}
Centuri\'{o}n M., Bonifacio P., Molaro P., \& Vladilo G. 2000, ApJ, 536, 540
\bibitem[2000]{Chen00}
Chen Y.Q., Nissen P.E., Zhao G., et al. 2000, A\&AS, 141, 491 
\bibitem[1997]{Chiappini97}
Chiappini C., Matteucci F., \& Gratton R. 1997, ApJ, 477, 765
\bibitem[1981]{Clegg81}
Clegg R.E.S., Lambert D.L., \& Tomkin J. 1981, ApJ, 250, 262
\bibitem[1998]{Dehnen98}
Dehnen W., \& Binney J.J. 1998, MNRAS, 298, 387
\bibitem[1997]{ESA97}
ESA, 1997, The Hipparcos and Tycho Catalogues, ESA SP-1200
\bibitem[1993]{Edvardsson93}
Edvardsson B., Andersen J., Gustafsson B., et al. 1993, A\&A, 275, 101
\bibitem[2000]{Finlator00}
Finlator K., Ivezic Z., Fan X., et al. 2000, AJ, 120, 2615
\bibitem[1987]{Francois87}
Fran\c{c}ois P. 1987, A\&A, 176, 294
\bibitem[1988]{Francois88}
Fran\c{c}ois P. 1988, A\&A, 195, 226
\bibitem[1998]{Fuhrmann98}
Fuhrmann K. 1998, A\&A, 338, 161
\bibitem[1973]{Garz73}
Garz T. 1973,  A\&A, 26, 471
\bibitem[2000]{Goswami00}
Goswami A., \& Prantzos N. 2000, A\&A, 359, 191
\bibitem[1998]{Grevesse98}
Grevesse N., \& Sauval A.J. 1998, Space Science Reviews, 85, 161
\bibitem[1992]{Hannaford92}
Hannaford P., Lowe R.M., Grevesse N., \& Noels A. 1992, A\&A, 259, 301
\bibitem[2000]{Idiart00}
Idiart T.P., \& Th\'evenin F. 2000, ApJ, 541, 207
\bibitem[1998]{Israelian98}
Israelian G., Garcia Lopez R., \& Rebolo R. 1998, ApJ, 507, 805
\bibitem[2001]{Israelian01a}
Israelian G., Rebolo R., Garcia Lopez R., et al. 2001, ApJ, 551, 833
\bibitem[2001]{Israelian01}
Israelian G. \& Rebolo R. 2001, ApJ, 557, L43
\bibitem[1987]{Johnson87}
Johnson D.R.H.,  \& Soderblom D.R. 1987, AJ, 93, 864  
\bibitem[2002]{Kurucz02}
Kurucz R.L. 2002, Atomic line list, available at his homepage (http://cfaku5.harvard.edu/linelists)
\bibitem[1984]{Kurucz84}
Kurucz R.L., Furenlid  I., Brault J., \& Testerman  L. 1984, Solar Flux Atlas from 296 to 1300nm, National Solar Observatory, Sunspot, New Mexico 
\bibitem[1994]{Nave94}
Nave G., Johansson S., Learner R.C.M., et al. 1994, ApJS, 94, 221
\bibitem[1992]{Nissen92}
Nissen P.E., \& Edvardsson B. 1992, A\&A, 261, 255
\bibitem[2001]{Nissen01}
Nissen P.E., Primas F., \& Asplund M. 2001, New Astronomy Reviews, 45, 545
\bibitem[2002]{Nissen02}
Nissen P.E., Asplund M., Chen Y.Q., \& Pettini M. 2002, in preparation
\bibitem[1983]{Olsen83}
Olsen E.H. 1983, A\&AS, 54, 55
\bibitem[1988]{Olsen88}
Olsen E.H. 1988, A\&A, 189, 173
\bibitem[1993]{Olsen93}
Olsen E.H. 1993, A\&A, 102, 89 
\bibitem[1996]{Ryan96}
Ryan S.G., Norris J.E., \& Beers T.C. 1996, ApJ, 471, 254
\bibitem[1998]{Samland98}
Samland M. 1998, ApJ, 496, 155
\bibitem[1989]{Schuster89}
Schuster W.J., \& Nissen P.E. 1989,  A\&A, 221,  65
\bibitem[2002]{Stephens02}
Stephens A., \& Boesgaard A.M. 2002, AJ, 123, 1647
\bibitem[2002]{Takada02}
Takada-Hidai M., Takeda Y., Sato S., et al. ApJ, in press, astro-ph/0103481
\bibitem[1999]{Thevenin99}
Th\'evenin F., \& Idiart T.P. 1999, ApJ, 521, 753
\bibitem[1995]{Timmes95}
Timmes F.X., Woosley S.E., \& Weaver T. 1995, ApJS, 98, 617
\bibitem[2000]{VandenBerg00}
VandenBerg D.A., Swenson F.J., Rogers F.J., Iglesias C.A., \&  Alexander D.R.
2000, ApJ, 532, 430 
\bibitem[1995]{Woosley95}
Woosley S.E., \& Weaver T.A. 1995, ApJS, 101, 181
\end{thebibliography}
\end{document}